\documentclass[conference]{IEEEtran}
\usepackage{cite}
\usepackage{xcolor,soul,framed} 
\colorlet{shadecolor}{yellow}
\usepackage{color,soul}
\usepackage[pdftex]{graphicx}
\graphicspath{{../pdf/}{../jpeg/}}
\DeclareGraphicsExtensions{.pdf,.jpeg,.png}
\usepackage[caption=false]{subfig}
\usepackage[cmex10]{amsmath}
\usepackage{algorithmic}
\usepackage{mdwmath}
\usepackage{mdwtab}
\usepackage{eqparbox}
\usepackage{url}
\usepackage{array}
\usepackage{fixltx2e}
\usepackage{dblfloatfix}
\usepackage[mathscr]{euscript}
\usepackage{amsfonts}
\usepackage{amsmath}
\usepackage{caption}
\usepackage{commath}
\usepackage{upgreek}
\usepackage[long]{optidef}
\usepackage{bm}
\usepackage{algorithm}

\usepackage{cleveref}
\usepackage{balance}
\usepackage{textgreek}
\usepackage{booktabs,chemformula}
\usepackage{array}
\usepackage{float}
\usepackage{amssymb}

\makeatletter
\def\footnoterule{\kern-3\p@
  \hrule \@width 2in \kern 2.6\p@} 
\makeatother
\begin{document}
\IEEEoverridecommandlockouts
\IEEEpubid{\begin{minipage}[t]{\textwidth}\ \\[35pt]
        \centering\small{978-1-7281-0270-2/19/\$31.00 \copyright2019 IEEE}
\end{minipage}} 

\title{Ray Tracing Analysis for UAV-assisted Integrated Access and Backhaul Millimeter Wave Networks}
\vspace{-5 in}
\author{\IEEEauthorblockN{Alberto Perez,
Abdurrahman Fouda,
and Ahmed S. Ibrahim
\thanks{This work is supported in part by the National Science Foundation under Award No. CNS-1618692.}}
\IEEEauthorblockA{Department of Electrical and Computer Engineering\\
Florida International University, Miami, Florida, USA 33174\\ Emails: \{apere669, afoud004, aibrahim\}@fiu.edu}} 

\maketitle

\begin{abstract}
The use of Millimeter-wave (mmWave) spectrum in cellular communications has recently attracted growing interest to support the expected massive increase in traffic demands. However, the high path-loss at mmWave frequencies poses severe challenges. In this paper, we analyze the potential coverage gains of using unmanned aerial vehicles (UAVs), as hovering relays, in integrated access and backhaul (IAB) mmWave cellular scenarios. Specifically, we utilize the WinProp software package, which employs ray tracing methodology, to study the propagation characteristics of outdoor mmWave channels at $\mathbf{30}$ and $\mathbf{60}$ GHz frequency bands in a Manhattan-like environment. In doing so, we propose the implementation of amplify-and-forward (AF) and decode-and-forward (DF) relaying mechanisms in the WinProp software. We show how the 3D deployment of UAVs can be defined based on the coverage ray tracing maps at access and backhaul links. Furthermore, we propose an adaptive UAV transmission power for the AF relaying. We demonstrate, with the aid of ray tracing simulations, the performance gains of the proposed relaying modes in terms of downlink coverage, and the received signal to interference and noise ratio (SINR).

\end{abstract}

\IEEEpeerreviewmaketitle

\section{Introduction}\label{sec_intro}

Next-generation cellular networks are expected to support ultra-high data rates to serve the exponential increase in capacity and traffic demands~\cite{5gtechs}. Millimeter-wave (mmWave) spectrum band is considered as a promising solution to achieve the required data rates, due to its large bandwidth ~\cite{tedmmwave, mmwavecellular}. However, transmission over mmWave band suffers from high propagation loss and low signal penetration, through buildings and solid materials, which results in regions with limited signal coverage~\cite{mmwaveref}.
To overcome such coverage gaps in mmWave communication, network densification, or the deployment of ultra-dense cellular networks (UDNs), has been recently proposed~\cite{UDCN,UDNmmWave}. However, the deployment cost and the spatio-temporal variability of capacity and coverage demands are among the challenges facing UDNs. 

Alternatively, integrating unmanned aerial vehicles (UAVs) into the cellular structure has been recently proposed (e.g.\cite{UAVhetnet,UAVRelay,OBIABUAV}), including the mmWave communication~\cite{uavmmwave}. Essentially, a UAV can relay the received data from the neighboring base station to its surrounding mobile users. Consequently, UAVs can be considered as a feasible, cost-effective and easily scalable network solution, which can be adjusted based on the coverage and capacity demands~\cite{totUAV}. Having a UAV communicating over backhaul links, towards base stations, and access links, towards mobile users, naturally leads to creating a wirelessly backhauled network architecture~\cite{UAVIBIAB}. 

In this regard, the integrated access and backhaul (IAB) network architecture is considered as a promising solution to allow for easier deployment of wirelessly backhauled networks. Generally, in an IAB architecture (e.g.~\cite{3GPPIAB,3GPPIAB_att}), the macro base station (MBS) uses the same infrastructure and wireless channel resources to provide access and backhauling functionalities for cellular users and IAB-relays, respectively. Despite the recent studies focusing on the coverage probability and interference mitigation in IAB 5G cellular networks (e.g.~\cite{BWPart, jointIBIAB}), there is no consideration of UAV-assisted IAB mmWave networks, which is the focus of this paper. 

UAV-assisted IAB mmWave networks may be studied via conducting \emph{ray tracing} simulations. Ray tracing approach has been recently utilized to characterize the mmWave signal propagation in indoor environments~\cite{ismailray, mmwaveindoor} and outdoor environments at both the 28 and 72 GHz bands~\cite{mmwaveray, airinterray}. However, none of the above studies has jointly studied the integration of UAVs into the mmWave-based IAB scenarios.

In this paper, we aim to characterize the coverage gains in UAV-assisted IAB mmWave network. More precisely, we utilize WinProp software package, which employs a ray tracing approach, to characterize the potential performance improvements of using UAVs as hovering relays in an IAB mmWave network. We focus on the outdoor mmWave channels at $30$ and $60$ GHz frequency bands. In doing so, we consider two relaying modes of UAVs, which are amplify-and-forward (AF) and decode-and-forward (DF). In the AF mode, we study the Out-of-Band IAB (OB-IAB) scenario and propose a power mapping table to determine the transmission power of UAVs at the access links, based on the received signal strength at the backhaul links. Furthermore, we define where to best position the UAVs, based on the generated ray tracing coverage maps. 

In the DF mode, we study the in-band IAB (IB-IAB) scenario and define the 3D deployment of UAVs. In doing so, we consider the required received signal-to-interference-plus-noise-ratio (SINR) threshold at the backhaul links along with the signal coverage and interference levels at the access links. To the best of our knowledge, this is the first ray tracing-based study that investigates the use of UAVs in mmWave-based OB and IB scenarios for IAB cellular networks.

The rest of this paper is organized as follows. In Section~\ref{sec_netarch}, we present the network architecture and the environment modeling in WinProp software. In Section~\ref{sec_AF}, we discuss the WinProp implementation of AF relaying mode and demonstrate, with the aid of the ray tracing simulations, the performance gains of using UAVs in IAB mmWave scenarios. Similarly, DF relaying mode is studied in Section~\ref{sec_DF}. Finally, conclusion is drawn in Section~\ref{sec_conc}. 

\section{Network Architecture and Ray-tracing Simulated Environment}\label{sec_netarch}

\begin{figure}
  \begin{center}
  \includegraphics[width=8.75cm,height=8.75cm,keepaspectratio]{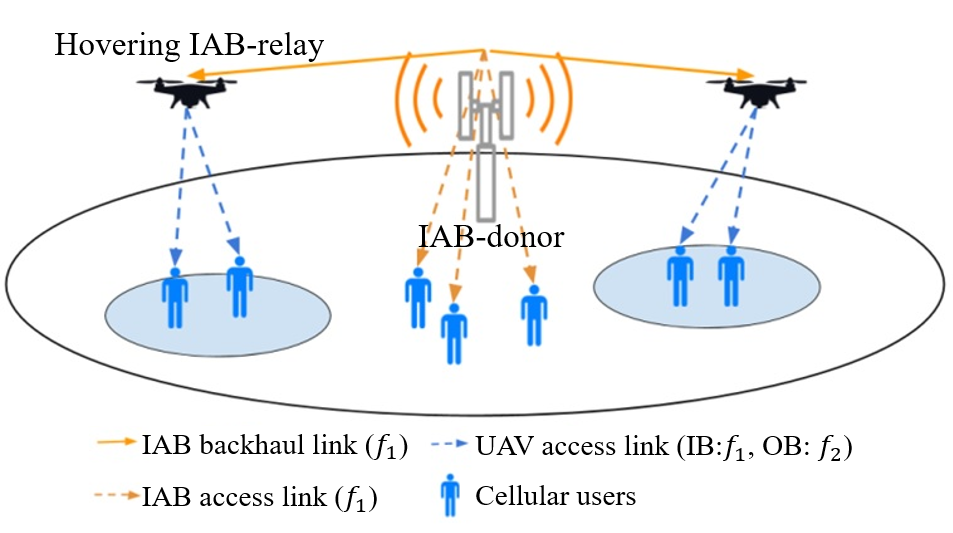}
  \caption{UAV-assisted integrated access and backhaul.}\label{fig_sysmod}
  \end{center}
  \vspace{-0.1 in}
\end{figure}

Fig.~\ref{fig_sysmod} depicts the considered architecture for the UAV-assisted IAB system.  
As shown, the IAB-donor, which is the MBS in this case, provides access and wireless backhauling functionalities to terrestrial users (tUEs) and UAVs, respectively. Both backhaul and access links operate on the same spectrum band, centered around the $f_1$ frequency. UAVs are utilized as relaying IAB-nodes to fill the coverage gaps and provide access functionality to aerial users (aUEs). The access links of UAVs operate at the same spectrum resources as backhaul links ($f_1$) in the IB-IAB transmission mode, and at different resources ($f_2$) in the OB-IAB transmission mode.

\begin{figure}
  \begin{center}
  \includegraphics[width=8.75cm,height=8.75cm,keepaspectratio]{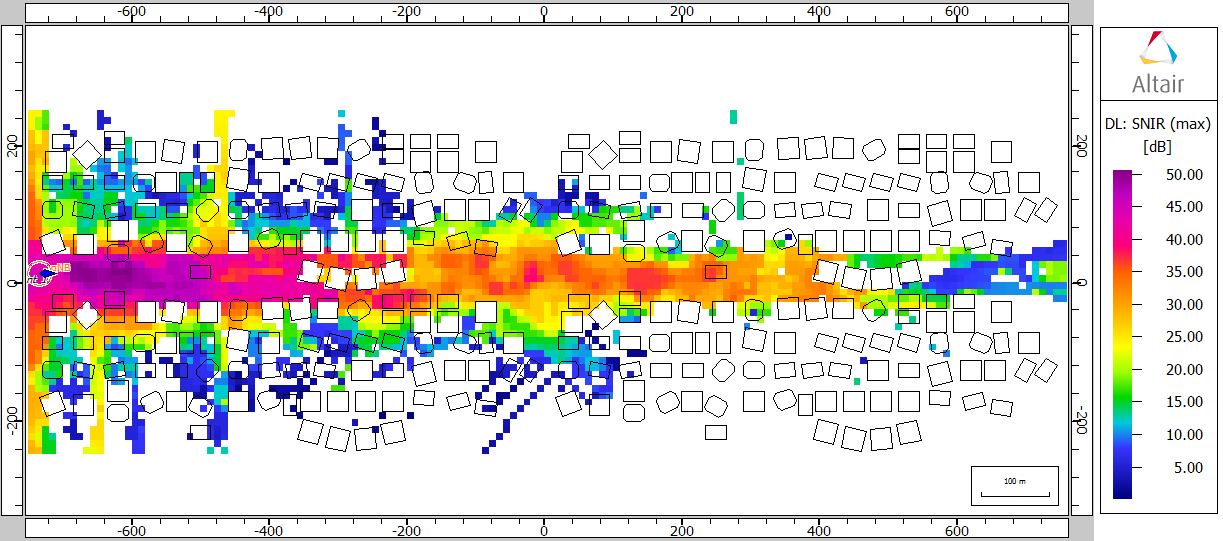}
  \caption{Received Downlink SINR from single IAB-donor.}\label{sc1_covrg_singleIABDon}
  \end{center}
    \vspace{-0.15 in}
\end{figure}
We utilize the WinProp software package to model a 3-dimensional (3D) urban outdoor scenario, namely the downtown Manhattan area, as shown in Fig.~\ref{sc1_covrg_singleIABDon}.
The 3D map layer allows for precise testing of the high path-loss limitations of mmWave spectrum bands. All structures are high rise buildings composed of concrete and flat glass surfaces. The heights of the buildings vary between $80\mathrm{m}$ and $120\mathrm{m}$. The buildings are uniformly distributed over a four-way intersection geographical area of size $1500\mathrm{m}\times460\mathrm{m}$. 
A single IAB-donor is positioned at the $(-700,0)$ coordinates at an altitude of $25\mathrm{m}$. 

\begin{figure}[h]
\vspace{-0.104 in}
    \begin{center}
        \subfloat[Azimuth plane radiation pattern\label{antennapatternA}]{
        \includegraphics[width=7.75cm,height=7.75cm,keepaspectratio]{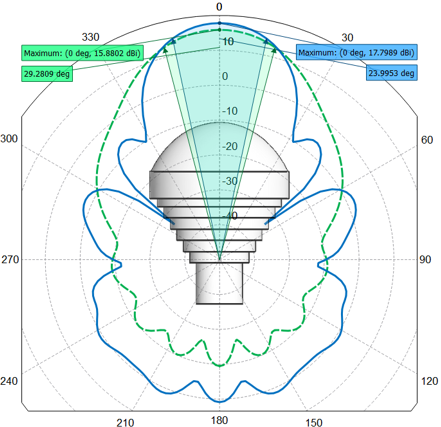}}
    \end{center}
    \begin{center}
        \subfloat[Elevation plane radiation pattern\label{antennapatternE}]{
        \includegraphics[width=7.75cm,height=7.75cm,keepaspectratio]{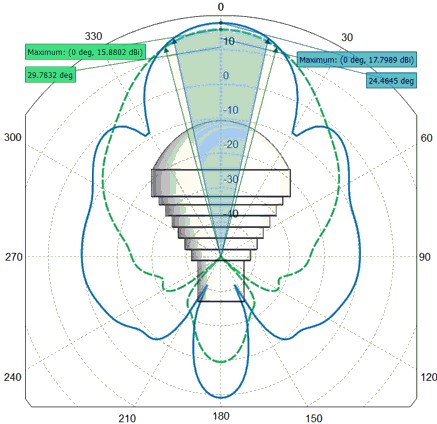}}
    \caption{Antenna radiation pattern.}
    \end{center}
    \vspace{-0.2 in}
\end{figure}

The high scattering nature of the simulated environment limits signal penetration and results in large coverage gaps. In this paper, we aim to shrink such coverage gaps by finding the best positions for UAVs. The UAV transmitter is implemented similar to a conventional transmitter with varying 3D location. The IAB-donor and UAVs utilize a directional antenna pattern for the ray tracing simulations. The 2D horn antenna radiation patterns are shown in Figs.~\ref{antennapatternA} and \ref{antennapatternE}. The antenna pattern has a beamwidth $\left(3~\mathrm{dB}\right)$ of ${\approx}~30$ degrees. The AMan module in WinProp software is utilized to post process these patterns and create a 3D antenna radiation pattern that is imported into the ray tracing simulations.

\section{Amplify-and-Forward Relaying in UAV-assisted OB-IAB mmWave Network}\label{sec_AF}
In this section, we address two design aspects related to the AF relaying mode. First, we propose a power mapping table for the AF relaying UAV, which defines its transmission power over the access link, based on its received power from the IAB-donor over the backhaul links. Second, we define the best locations for the group of UAVs to be deployed.    


\subsection{AF Power Mapping and UAV's Positioning}\label{subsec_AFpwrmap}
The received SINR at the backhaul link of a UAV depends on its location and distance from the IAB-donor. Hence, we propose an adaptive power transmission scheme, to be utilized at the AF UAVs, to reflect the strength of the backhaul connection of a UAV. In other words, the higher the received power over the backhaul links, the more transmission power will go over the access links. 
\begin{figure}
  \begin{center}
  \includegraphics[width=8cm,height=8cm,keepaspectratio]{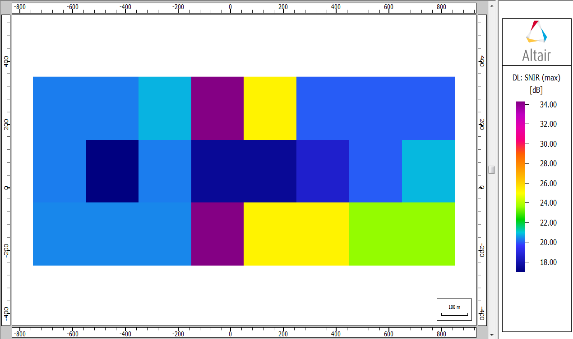}
  \caption{Received SINR levels at backhaul link of a single UAV.}\label{sc1_sinr_bh}
  \end{center}
      \vspace{-0.15 in}
\end{figure}

First, we conduct ray tracing simulation to generate a power map of the received signal strength over the backhaul link at the UAV altitude of $200$~m. Given that this is a single-transmitter simulation environment, there is no interference and the signal strength and SINR are proportional to each other. 
Fig.~\ref{sc1_sinr_bh} depicts a sample coverage map with a resolution of $200$m of the received SINR levels at the UAV from a single IAB-donor. Each square represents a potential UAV location. 

Second and for a generic UAV's location, $i$, represented by one of the squares in Fig.~\ref{sc1_sinr_bh}, we propose to utilize a UAV transmission power, $P_{i}^{(\mathrm{Tx})}$, equal to  
\begin{equation}
P_{i}^{(\mathrm{Tx})}=P^{(\mathrm{max})}\times\left(\frac{\upgamma_{i}^{(\mathrm{BH})}}{\upgamma_{\mathrm{u}}^{(\mathrm{max})}}\right) \;, 	    
\end{equation}
where $P^{(\mathrm{max})}$ and $\upgamma_{i}^{(\mathrm{BH})}$ denote the maximum power capability of the UAV and the received SINR value at the $i^{\mathrm{th}}$ potential UAV location. Furthermore, $\upgamma_{\mathrm{u}}^{(\mathrm{max})}$ represents a normalization factor, which is equal to the maximum received SINR at the user's altitude level. 

\begin{table}[h]
  \centering
  \caption{AF UAV power mapping table.}\label{tab_pwrmap}
        \begin{tabular}{|c|c||c|c|}
        \hline
        Received  & UAV Tx & Received  & UAV Tx\\
        SINR (dB) &  power (dBm) &  SINR (dB) &  power (dBm)\\
        \hline
        16.95 & 32.28 & 20.37 & 33.10\\
        \hline
        17.42 &	32.41 &	20.74 &	33.16\\
        \hline
        17.45 &	32.43 &	20.81 &	33.18\\
        \hline
        18.56 &	32.67 &	23.47 &	33.69\\
        \hline
        19.98 &	32.99 &	25.17 &	34.00\\
        \hline
        20 & 33.01 & 25.21 & 34.01\\
        \hline
        20.24 &	33.05 &	34.22 &	34.42\\
        \hline
        20.27 &	33.07 &	34.32 &	34.43\\
        \hline
        \end{tabular}
\end{table}

Table~\ref{tab_pwrmap} shows non-repeated entries of the received SINR at the backhaul link obtained from the coverage map, in Fig.~\ref{sc1_sinr_bh}, and the corresponding transmission power calculated according to (1). Third, the best position for the UAV is determined based on 1) its received SINR level from the IAB-donor, as shown previously in Fig.~\ref{sc1_sinr_bh}, and 2) its potential impact on the coverage map at the ground user level, taking into consideration the power mapping table presented in Table~\ref{tab_pwrmap}. Any additional UAV can be deployed similarly in a consecutive manner.    



\subsection{Coverage Enhancement}\label{subsec_AFsimres}

\begin{figure}
  \begin{center}
  \includegraphics[width=8.5cm,height=8.5cm,keepaspectratio]{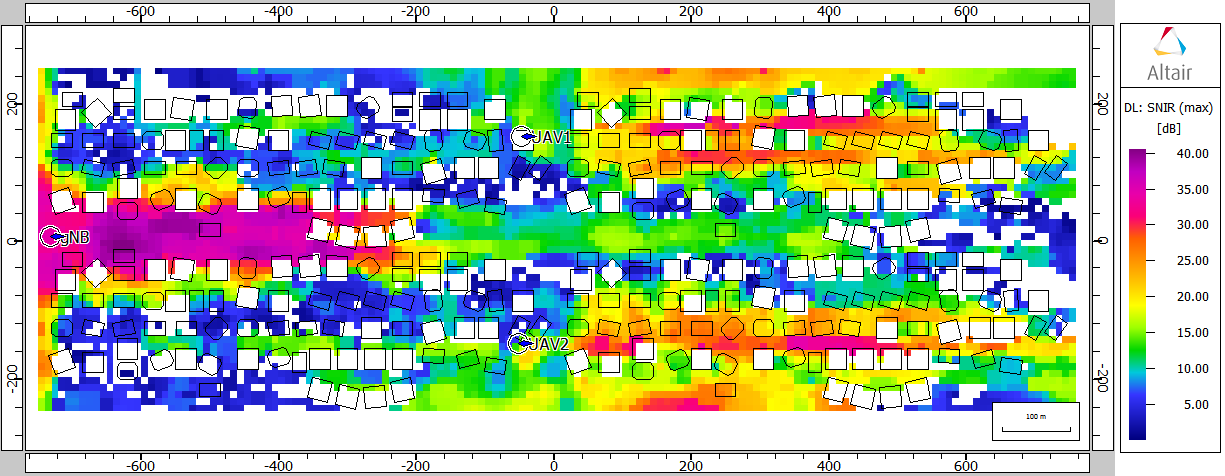}
  \caption{Ray tracing coverage map after adding two AF UAVs.}\label{sc1_covg_2uavs}
  \end{center}
  \vspace{-0.15 in}
\end{figure}

In this section, we show the coverage improvement due to deploying two UAVs in the UAV-assisted OB-IAB mmWave network. We consider the same baseline deployment scenario, shown previously in Fig.~\ref{sc1_covrg_singleIABDon}, in which a single IAB-donor is positioned at the $(-700,0)$ position and at an altitude of $25\mathrm{m}$. In the considered UAV-assisted OB-IAB scenario, $30$~GHz frequency band is used for the downlink transmissions of backhaul and access links of the IAB-donor. Given its \emph{out-of-band} nature, the access links of UAVs operate at a different frequency, which is the $60$~GHz frequency band. The IAB-donor and any UAV has a maximum downlink transmission power of $10$ and $5$ watts, respectively. Based on the UAV positioning and transmission power mapping approach, introduced in section~\ref{subsec_AFpwrmap}, the best positions for the two UAVs are found to be $(-50, 150)$ and $(-50, -150)$. The two UAVs are positioned almost at the middle of a four-way intersection at an altitude of $200$m. The deployed UAVs leverage their Line-of-Sight (LOS) capabilities and fill the coverage gaps as shown in the received downlink SINR coverage map in Fig.~\ref{sc1_covg_2uavs}.

\begin{figure}
  \begin{center}
  \hspace{-.45in}
  \includegraphics[width=8.5cm,height=8.5cm,keepaspectratio]{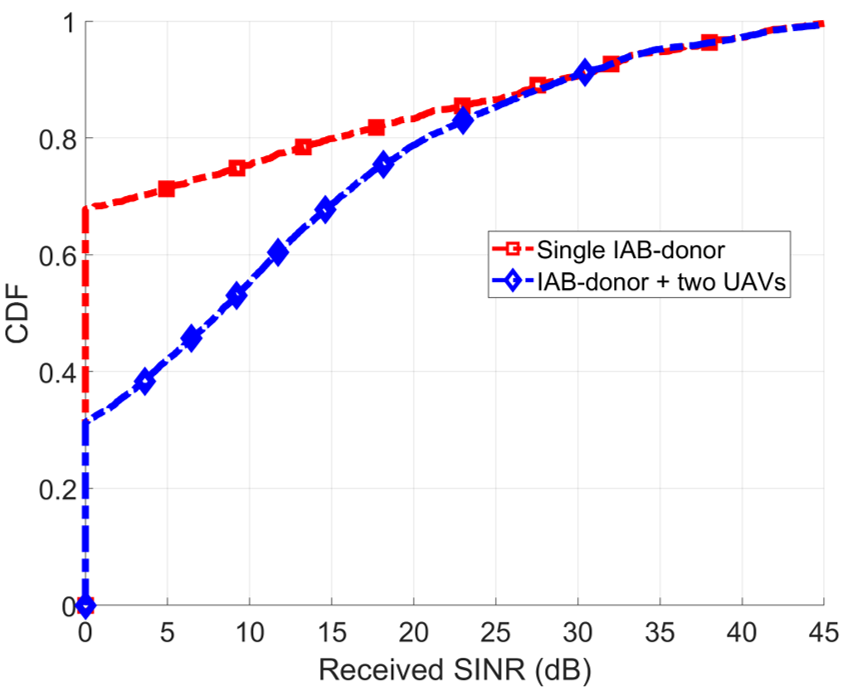}
  \caption{AF relaying mode: CDF of downlink received SINR.}\label{sc1_cdf}
  \end{center}
    \vspace{-0.1 in}
\end{figure}

Fig.~\ref{sc1_cdf} depicts the cumulative distribution function (CDF) of the received downlink SINR of the baseline scenario, in which UAVs are not used, and the UAV-assisted scenario. In both cases, the users are distributed uniformly across the coverage map every $10$~m. Fig.~\ref{sc1_cdf} shows that initially around $70^{\mathrm{th}}$ percentile of users fall in coverage gaps in the baseline scenario. With the deployment of two UAVs, only $30^{\mathrm{th}}$ percentile are still left in coverage gaps. In other words, deploying two UAVs, according to the proposed scheme in this paper, has provided an average of $2.3\times$ gain in downlink coverage.

\section{UAV-Assisted Decode-and-Forward Relaying in UAV-Assisted IB-IAB mmWave Network}\label{sec_DF}
\begin{figure}
  \begin{center}
  \includegraphics[width=8.75cm,height=8.75cm,keepaspectratio]{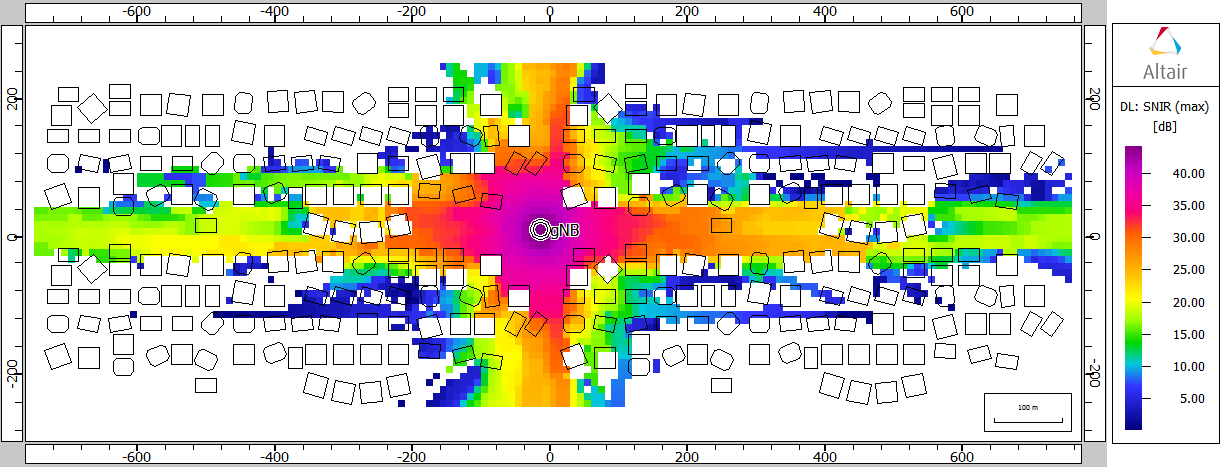}
  \caption{Received downlink SINR map at user's altitude.}\label{sc2_covg_single_IABDon}
  \vspace{-0.15 in}
  \end{center}
\end{figure}

In this section, we consider DF relaying nature of the UAVs, and aim to define the best locations of a set of deployed UAVs. In the DF relaying mode, a UAV forwards its received packet, if the received SINR is above a certain threshold. Otherwise, it remains idle. The transmitted signal will be sent with the maximum transmission power of the UAV. In this section, we use IB-IAB transmission mode, as a potential candidate for tighter integration between access and backhaul links. In that, the backhaul and access links of each UAV fully overlap on spectrum resources. The baseline simulation scenario is depicted in Fig.~\ref{sc2_covg_single_IABDon}, in which, the IAB-donor is positioned at the origin coordinates with an altitude of $25\mathrm{m}$. 
 
\subsection{DF UAVs Positioning}\label{subsec_DFimp}
\begin{figure}
  \begin{center}
  \includegraphics[width=8.75cm,height=8.75cm,keepaspectratio]{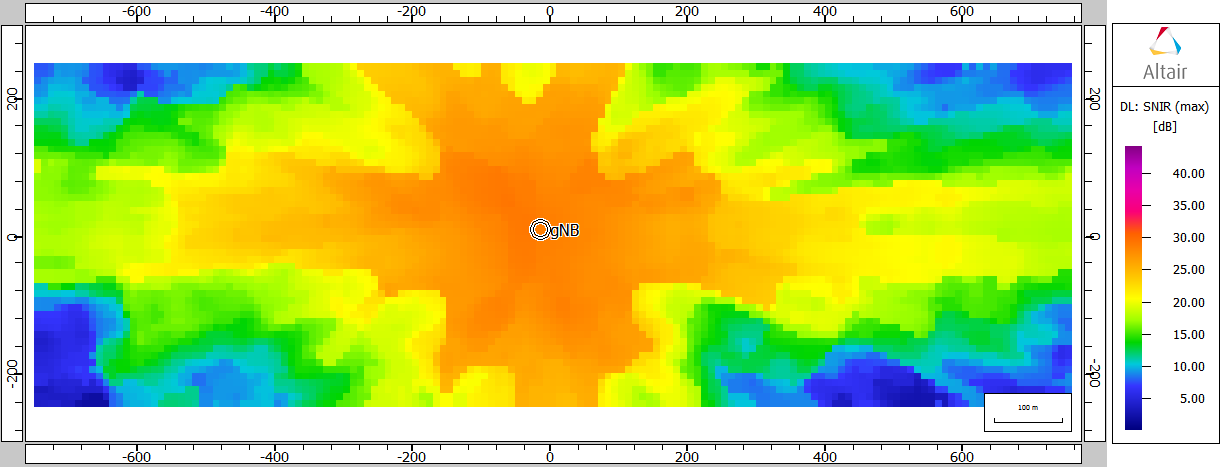}
  \caption{Received downlink SINR map at backhaul link of a single UAV.}\label{sc2_covg_bh}
  \end{center}
    \vspace{-0.15 in}
\end{figure}

We utilize two UAVs to fill in the coverage gaps in Fig.~\ref{sc2_covg_single_IABDon}, taking into consideration the inter-UAV interference to be at low levels. In particular, we conduct ray tracing simulation to generate the coverage map of the received SINR at the backhaul links, as depicted in Fig.~\ref{sc2_covg_bh}. On one hand, each UAV must be positioned in a location where the received backhaul SINR from the IAB-donor is above a specific threshold. On the other hand, the UAVs are positioned such that the signal coverage is maximized at the ground user level while taking into account minimizing the inter-cell interference levels. We set the SINR threshold to $15\,\mathrm{dB}$ in the proposed ray tracing scenario. Hence, the UAVs can be placed anywhere where the received SINR in Fig.~\ref{sc2_covg_bh} is above this threshold. The best positions for the two UAVs are found to be $(-20,200)$ and $(20,-200)$.


\subsection{Coverage Enhancement}\label{subsec_DFres}
\begin{figure}
  \begin{center}
  \includegraphics[width=8.75cm,height=8.75cm,keepaspectratio]{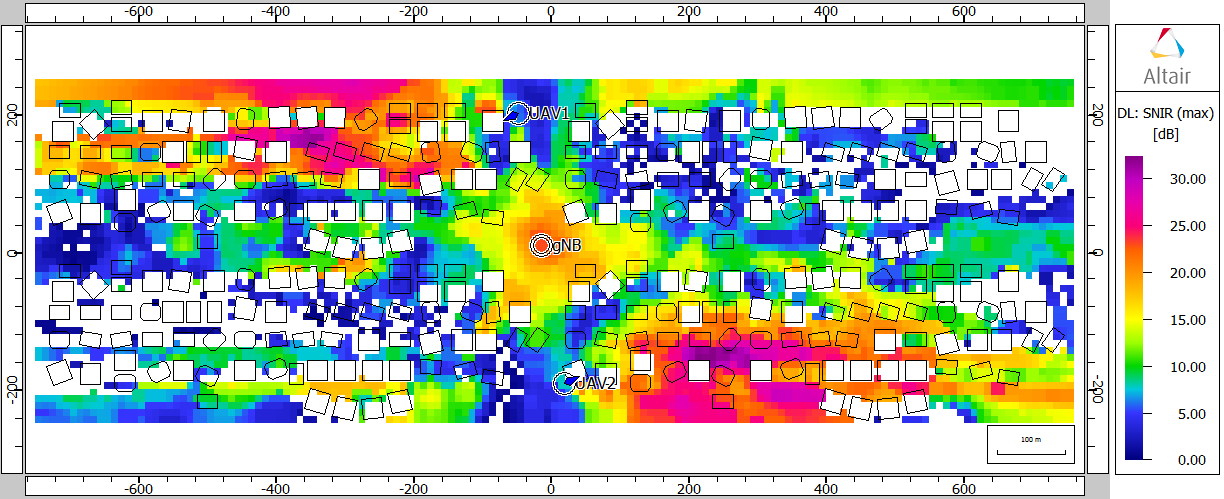}
  \caption{Ray tracing coverage map after adding two DF UAVs.}\label{sc2_covg_2uavs}
  \end{center}
    \vspace{-0.15 in}
\end{figure}
\begin{figure}
  \begin{center}
  \hspace{-.45in}
  \includegraphics[width=8.5cm,height=8.5cm,keepaspectratio]{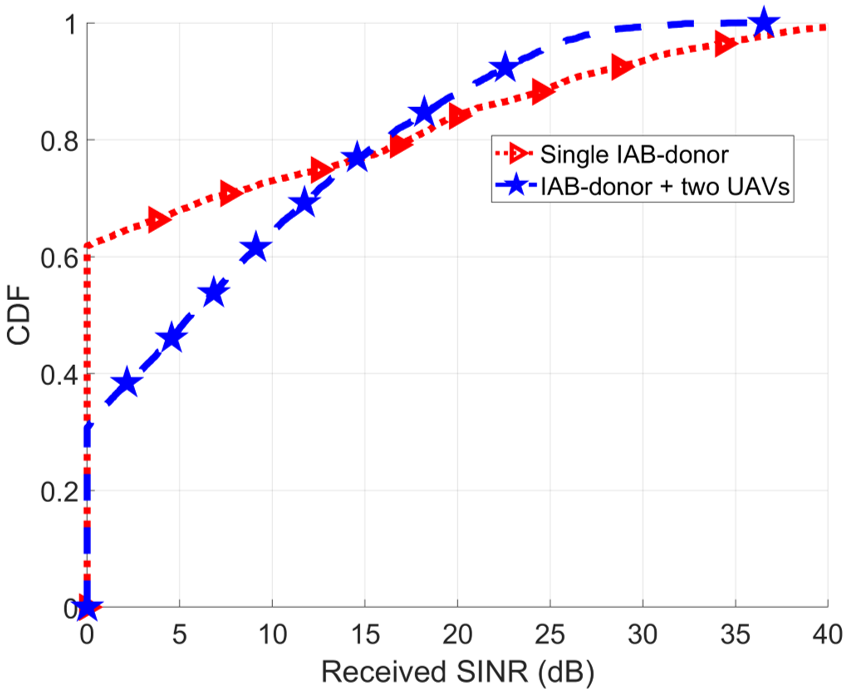}
  \caption{DF relaying mode: CDF of downlink received SINR.}\label{sc2_cdf}
  \end{center}
  \vspace{-0.15 in}
\end{figure}

Fig.~\ref{sc2_covg_2uavs} shows how UAVs are positioned to fill the coverage gaps in the baseline scenario (see Fig.~\ref{sc1_covg_2uavs}) and demonstrates the improvement in the downlink coverage after deploying the UAVs. The CDF plot of the received downlink SINR is shown in Fig.~\ref{sc2_cdf}. Fig.~\ref{sc2_cdf} shows that $60^{\mathrm{th}}$ percentile of users suffer from coverage gaps before deploying the UAVs, while only $30^{\mathrm{th}}$ percentile do after deploying the UAVs. In other words, the use of two DF UAVs yields an average of $1.75\times$ gain in downlink coverage of the proposed IAB mmWave scenario. It is worth noting that the available spectrum resources are directly proportional to the number of UAVs in the DF relaying mode. Consequently, given that deploying two UAVs has doubled the downlink coverage, Fig.~\ref{sc2_cdf} shows that the downlink capacity achieves $2\times$ gain in the proposed DF relaying mode. Fig.~\ref{sc2_cdf} also reveals that the received downlink SINR of terrestrial users, i.e., cell-center users, is slightly decreased by around $4\,\mathrm{dB}$ to provide coverage to more than $30^{\mathrm{th}}$ percentile of users. This degradation is due to the nature of \emph{in-band} mode, which creates mutual interference between the access links of the UAVs and the IAB-donor.  

\section{Conclusion}\label{sec_conc}
In this paper, we utilized the ray tracing simulations in WinProp software to investigate the coverage gains of using UAVs, as hovering relays, in IAB mmWave cellular networks. We considered the AF and DF relaying modes of UAVs and analyzed the propagation characteristics of $30$ and $60$ GHz outdoor channels. We used the ray tracing simulation results to define the 3D deployment and the access functionality of the UAVs. The ray tracing simulation results show that using UAV AF and DF relaying modes achieves an average of $2.3\times$ and $1.75\times$ gains in the downlink coverage of IAB mmWave networks, respectively.

\balance
\bibliographystyle{IEEEtran}
\bibliography{main2,main}
\end{document}